# Relevance Judgment Convergence Degree – A Measure of Inconsistency among Assessors for Information Retrieval

**Dengya Zhu**
*School of Management, Curtin University*
*Perth, WA, Australia*                              *d.zhu@curtin.edu.au*

**Shastri L Nimmagadda**
*School of Management, Curtin University*
*Perth, WA, Australia*                 *shastri.nimmagadda@curtin.edu.au*

**Kok Wai Wong**
*Murdoch University*
*Perth, WA, Australia*                        *k.wong@murdoch.edu.au*

**Torsten Reiners**
*School of Management, Curtin University*
*Perth, WA, Australia*                          *t.reiners@curtin.edu.au*

## Abstract

Relevance judgment of human assessors is inherently subjective and dynamic when evaluation datasets are created for Information Retrieval (IR) systems. However, a small group of experts' relevance judgment results are usually taken as ground truth to "objectively" evaluate the performance of the IR systems. Recent trends intend to employ a group of judges, such as outsourcing, to alleviate the potentially biased judgment results stemmed from using only a single expert's judgment. Nevertheless, different judges may have different opinions and may not agree with each other, and the inconsistency in human relevance judgment may affect the IR system evaluation results. In this research, we introduce a Relevance Judgment Convergence Degree (RJCD) to measure the quality of queries in the evaluation datasets. Experimental results reveal a strong correlation coefficient between the proposed RJCD score and the performance differences between the two IR systems.

**Keywords:** relevance judgment, information retrieval, information system development.

## 1. Introduction

Searching for information from the Internet is becoming a ubiquitous activity that is part of our daily life for all sections of society. As highlighted in Saracevic [24], most people seek information according to what they know at hand, and assume that the relevance is based on a "y'know" notion that does not require explanation. The new of information relevance is also adjudged for the success of search engines, while a formal, complex definition of such relevance is presented in [24]. Furthermore, one of the basic objectives of any IR system is to improve the relevance of search results after a user query (used to express the user's information need) is submitted to an IR system. This basic objective is not limited to IR systems, but also includes recommendation systems, advertising systems, and scientific database systems. Consequently, appropriate measurement criteria of search queries and ground truth test dataset are essential in Information System (IS) development.

The user – IR system interaction pattern implies two sets of relevance. Firstly, the IR "system relevance" is the returned search results that are believed relevant to what the user expressed in the queries that submitted to IR systems. Secondly, the "human relevance" is what an IR system user is looking for. The two sets of relevance may overlap perfectly; and on the other hand, the returned results believed by an IR system that are relevant may not match users' relevance. Therefore, in most cases, the user or social (a group of users) relevance judgment is usually taken as a gold standard for comparison [11, 22, 24].

While different IR systems may return different results for the same query, the relevance judgments made by different users vary as well [3, 6, 10, 14, 29]. Bailey et al. [3] grouped assessors into three categories as "gold", "silver" and "bronze" standards. They found relevant judgment agreements in terms of Jaccard Coefficient among the groups were relatively low, which is only about a third agreed with each other. Consequently, the low level of agreement among assessors negative impacted the construction of testing datasets used to estimate the performance of different IR systems [3]. In addition, relevance judgment results by "gold standard" group, who were experts in specific information retrieval tasks, usually deliver consistently better outcomes when the performance of an IR system is a concern, rather than just ranking the IR systems [3, 29].

Expert-based relevance judgment and IR system evaluation approach is expensive. As indicated in [1, 24], the evaluation requires a list of resources such as infrastructure, money, time, and organization; and it cannot be scaled up easily. To address the issues, a crowdsourcing approach using Amazon's Mechanical Turk is proposed to answer, "Can we get rid of TREC[1] assessors?". As described in [1], Jaccard Coefficient, Cohen's Kappa and its variants such as Fleiss's Kappa and Krippendorff's Kappa are used to measure the agreement among assessors based on the type of assessments. The researchers found that the agreement degree between crowdsourcing works and TREC experts could reach about 70% when crowdsourcing results are grouped and averaged. The crowdsourcing results could be considered reliable in the interpretation done in [1].

The aforementioned research assumed that 1) full text documents are available so judges can read the full document before making a relevant decision such as the TREC evaluation. While currently, most of us need to make a relevancy judgment simply based on only Web search results (usually less than 30 words), or Web snippets which are nevertheless more challenging compared to the availability of complete documents. Of course, a user can always click the link in the Web search result to read the full text. Nevertheless, our assumption is that one of the goals of a search engine is to improve search relevance, and the Web snippet itself should present sufficient concise information to facilitate users to make relevant judgments effectively, rather than let users click each link to read the full text to make relevance judgments. 2) experts' relevance judgments are taken as ground truth, even if there were disagreements between TREC experts and other judgement groups [1, 2, 5], and thus judging the quality of the test dataset is questionable. This may result in the evaluation results, either in terms of performance (with regard to precision and recall) or the ranking of the IR systems, were justified on a biased dataset. 3) Almost all the research so far focuses on measuring disagreements among assessors, and few studies on the quality of the queries, especially in Web IR systems evaluation.

The contributions of the research are 1) RJCD is proposed as a novel criterion to measure the quality of a query for Web IR systems, rather than the disagreements among assessors or assessor groups which are what other researchers have conducted by using Kappa and its variants; 2) in our experiments, we introduced "no sufficient information to make a decision" item to reduce the coincidental of randomly guesses of assessors, and thus further facilitate to improve the quality of test datasets.

## 2.   Related Work

Relevance can be defined as "the ability (of an information retrieval system) to retrieve material that satisfies the needs of the user" [21]. The concept of relevance has been well known in the area of information retrieval since the late 1950s. According to different assumptions, relevance can be categorized as system-oriented and user-oriented [9, 11, 12, 22, 24, 25], and thus relevance judgment for the two different types of relevance varies on different criteria. A more complex definition of relevance described by Mizzaro [17] was that relevance is a point in a four-dimensional space named as information resources, user

---

information needs representations, time and component such as task and topic, and context. Based on this definition, relevance judgment is somewhat "relevance indetermination on phenomenon" meaning that measuring users' real relevance is difficult. Hjørland [11] reviewed Saracevic's perspective of relevance is fundamentally a "subject based knowledge view" [22], and argued that "the user view" in [22] could be extended to a social point of view, that is, a purely individual view of relevance should not be used as the guideline for designing information systems and services; instead, a consensus view of relevance is more important in practice. Hjørland [11] further found that judges had difficulty to distinguishing relevant documents from irrelevant ones; Saracevic [23] provided a detailed review of inconsistent relevance judgment issues. The review work however has no clear suggestion on how the issues are resolvable.

Identifying objects or entities, which can take various forms such as documents, images, music, audio, and video, that are of interest and relevance to users' information need is the critical issue in the areas of IR and IS [24]. Among all the challenges in IR and IS, as argued by [11], the notion of relevance is implicated by not only IR itself but also involves cognitive science, logic, philosophy, and domain oriented [11]. As early as 1975, Saracevic [20] discussed five relevance models: system's view, user's view, subject literature view, subject knowledge view, and pragmatic view; however only the first two are widely cited. The system's view implies how relevance is viewed technologically and algorithmically; while the user's view can be described as "the subject knowledge view" which is believed as the most fundamental perspective of relevance [11]. However, Hjørland [11] further pointed out that, biased, individual/idiosyncratic relevance judgment is problematic if used as guidelines for information system development.

Agreement among judges is one of the subjective aspects of relevance in Mizzaro's model [16]. One concern is how the inconsistency of human relevance decision affects the IR evaluation results. Saracevic [24] found that until 2016, there were only seven studies that addressed the issue. Voorhees [29] mentioned that although a consistent conclusion is the inconsistency of assessors seems have only marginal effects on the relative performance of the evaluated IR systems, the averaging policy hides the performance of a given query, thus a limitation yet to be addressed.

To diminish the subject knowledge view of assessors, for TREC evaluation collection, experimental results of [26] revealed that randomly selected "relevant documents" from pooled documents (system's view of relevance) can also exhibit the ability to keep the same performance ranking order of IR systems. This approach has been further developed by [30] where each pooled document is assigned a reference score, and the accumulated scores of different IR systems are compared to decide the rank list of the systems. Spearman and Kendall Tau correlation coefficients are used to compare their ranked list with the official TREC ranked list.

In an interactive IR environment, relevance feedback and automatic query expansion are enabled. In addition, relevance is extended from dichotomous bipolar to highly relevant, fairly relevant, marginally relevant, and irrelevant. Experimental results from 26 participants with the TREC dataset demonstrated users can identify the most highly relevant and half marginal relevant documents [28]. At the same time, users may select off-topic documents for relevance feedback, and thus making the reliability of the relevance feedback results of users questionable. Data topicality judgement, data reliability judgement and data utility judgement patterns are identified which further benefited the designs of cognitive retrieval systems. Various preferences, scores and ties are used to analyze relevance judgements by comparing relevant scales [27].

To address the expensive, time consuming, assessor error, and potential disagreement issues in relevancy judgment, crowdsourcing approaches have been proposed to label the relevance of a test document set [1, 7, 13, 19]. Experiments demonstrated the assessor errors and tasks or domain knowledge of assessors are all factors that can affect the final IR systems performance ranking results [7]. By assigning the same judgment work to five outsourcing assessors, the binary relevance judgment agreement between TREC and the averaged results (three out of five) of outsourcing assessors are 77%. In case of disagreement, outsourcing results are more reliable [1]. However, an individual agreement between the two groups is relatively low, with Fleiss's Kapa only 0.195. To control the quality of outsourcing relevance judgment, the following factors need to be considered: how workers choose the topic of interest; how requesters can find quality workers and their knowledge areas; how to scale up outsourcing and keep quality; and various methods to estimate correlations between TREC experts and outsourcing workers [19].

## 3.   Issues and Challenges of Information Relevance Judgement

The ambiguous character of natural language and subjective feature of relevance judgment are the sources of the issues. Briefly, they are 1) challenging for search engines to return relevant search results for a given ambiguous search term, and 2) expensive to obtain sufficient labelled training data [1, 19, 24] for supervised learning; and consequently, the readily labelled training datasets are surprisingly scarce [8]. Further, labelling a document involves relevance judgments by human experts. In contrast, the objectiveness of relevance judgment as per categorization is an arguable topic [4, 11, 16, 17, 21, 22] as the relevance judgment itself is a subjective outcome. In addition, both supervised and unsupervised machine learning algorithms are developed and evaluated based on full-length text documents [15, 32]. However, for Web IR, the text to be processed, either manually or automatically by using machine learning algorithms, is the Web snippet, which is less informative than the full-length text and is very sensitive to how the Web snippets are algorithmically extracted and presented by different Web IR systems [32].

The less informative aspects of Web search results have significant implications for relevance judgments which is the core of IR [17, 21]. Without prior domain knowledge, adaptation of interpreted search results affects the relevance judgement discernment and its inference in user preferences and scores.

Another issue is that we take user relevance judgment results as ground truth to evaluate our IR algorithms. If the relevance judgment is seriously subjective towards only personal preferences biased judgments, the evaluated results can hardly be used as an objective measurement of the performance of the developed algorithms or Web IR systems.

## 4.   Motivation and Research Goal

Motivation and reasoning capacity are vital variables in major social judgment and persuasion models. Literature suggests cognitive performance has high level of motivation that may be detrimental to information judgement performance, mainly when cognitive resources are rare. Test collection is a critical motivation in evaluating the information retrieval systems. Generating relevance judgements involves expensive and time-consuming human assessors. These issues have motivated us to adopt innovative and inexpensive crowdsourcing method for data acquisition. For accuracy and reliability of judgements, the current research is the motivation.

Information systems in the contexts of the interpretation of Web search results construe two focused elements: IR and storage modelling. The current research adds another element "interpretation" to adjudge the information relevance judgement in developing effective retrieval or Web IR systems. The goals of our research are to 1) develop a measuring mechanism of RJCD that can be used to create a less subjective ground truth test dataset for evaluating Web IR systems and algorithms; 2) verify the proposed RJCD has a positive coefficient with the improvement of a Web search results classification and re-rank model; 3) use open-source experimental data, including search queries and the corresponding information needs. Sample questionnaires and all search results with the queries are at https://github.com/simon-oz/relevance-judgement.git.

## 5.   Research Methodology and Web Search Instrument Development

Our research intends to address the above issues which are emerged as a central notion in information science development but have not yet attracted sufficient attention. First, we created a dataset using the search results from a meta search engine. This curated dataset can alleviate the knowledge of relevance judgment inconsistently as discussed in [24]. Since Web snippets may not contain enough information to make a judgment, we introduced a "no sufficient information to make a decision" option to avoid potential random guesses of judges. Second, we introduced the Relevance Judgment Convergence Degree (RJCD) as a measurement to decide if a data item should be included in the testing dataset to maintain the quality of the ground truth dataset. To validate the proposed approach, we have conducted experiments by comparing our re-ranked results with that of a meta-search engine and presented our experimental results that illustrate the positive coefficient between JDC and performance improvement in terms of precision.

The research aims to evaluate how the proposed RJCD can be employed to address the subjectiveness issue of IR system evaluations. Exploratory and descriptive with empirical research are used to describe and interpret different terminologies with the instances of relevance judgements. In the empirical research, we compare search results improvement between a meta-search engine which uses Yahoo Search Web Service APIs, and a Web search results re-ranking model which classifies the search results into top level topics of the Open Directory Project (ODP)[2]. We further re-rank the classified results based on user preference profile [33]. Jansen and Spink [12] found that most users browse only several results in pages, and more than half of the users view only the first page returned by search engines. Therefore, we limit only the top 50 returned items from our meta search engine for each of the 30 queries as discussed in the following sections; and the returned items are then categorized into different ODP categories.

## 6. Ambiguous Search term Selection and Relevance Judgment

### 6.1. Ambiguous Search term Selection

Search terms used to evaluate IR systems play a critical role because different IR systems usually return different search results for the same information needed when expressed as search terms. Traditionally, the performance of an IR systems is evaluated by a relatively small human labelled dataset such as TREC with predefined search terms; and an IR system is expected to return as many known relevant documents and as few known irrelevant documents as possible. In the age of information explosion, especially in the area of Web search, search terms submitted to a search engine are different from the well-predefined search terms, as Web users are not limited to only academic staff when TREC was designed ; but include people with various educational backgrounds and knowledge areas.

Therefore, the following principles [33] are employed as a guideline to select search terms which are used in our experiments to evaluate the performances of a baseline IR system and a re-ranking system.

1) Real search terms from real users.
2) Search terms are short and contain only one or two words.
3) Search terms should cover a variety of topics.

Researchers have suggested the minimum number of queries when evaluating an IR system. Zeng et al. [32] used 30 queries with 200 top ranked search results to evaluate the performance of three search engines: Alta Vista, MSN and Google. Manning et al. [15] believe 50 queries is the minimum number for IR evaluation. Buckley and Voorhees [5] suggested that a good experiment needs 25 to 50 queries to produce the desired confidence in experimental results. Xu and Chen [31] found and suggested more search terms would generate more reliable conclusion about the performance of an IR system. Nevertheless, generation of ground truth datasets used to estimate IR systems requires expensive human experts to label the dataset by judging if a document is relevant or not to a given query. Human relevancy judgments per se are inherently subjective which may result in a biased ground truth dataset and scaling up the dataset is empirically difficult [9, 14, 25]. Considering the human cost, scale of the experiments in the research and without losing significance of the experiment results, we selected 30 queries as listed in Table 1. All the

---

search terms are real user search terms submitted to the Microsoft MSN search engine [32] wiht three categories, "Ambiguous terms", "Entity names" and "General terms" [12].

Table 2 shows the statistical information of the 30 search queries. Among the 30 queries, 84% (25/30) have single work queries, 13% (4/30) have two-work, and 3% have three-word queries.

**Table 1.** Queries used in experiments [33]

| Search term | | Your information need |
|---|---|---|
| **Ambiguous Terms** | apple | apple computer company |
| | jaguar | animal jaguar |
| | saturn | the planet Saturn |
| | jobs | the person Steve Jobes |
| | jordan | the Hashemite kingdom Jordan |
| | tiger | the animal tiger |
| | trec | Text Retrieval Conference |
| | ups | the Uninterrupted Power Supply |
| | quotes | how to correctly use quotes in writing |
| | matrix | the mathematics concept matrix |
| **Entity names** | susan dumais | the researcher Susan Dumais |
| | clinton | the US ex-president, Bill Clinton |
| | iraq | general geographic and demographical information about iraq |
| | dell | the dell computer company |
| | disney | the person Walt Disney |
| | world war 2 | history related to world war 2 |
| | ford | Henry Ford, the founder of the Ford Motor Company |
| **General terms** | health | how to keep healthy |
| | yellow pages | the origin of yellow pages |
| | maps | how to read maps |
| | flower | wild flower |
| | music | music classification by Genre |
| | chat | computer-mediated chat systems |
| | games | history of games |
| | radio | history of radio |
| | jokes | the most funny jokes |
| | graphic design | the art and practice of graphical design |
| | resume | how to write a resume |
| | time zones | time zones of the world |
| | travel | travel planning and preparation |

**Table 2.** Features of search terms

| Categories | Single term | Two terms | Three terms | Total |
|---|---|---|---|---|
| **Ambiguous terms** | 10 | | | 10 |
| **Entity names** | 5 | 1 | 1 | 7 |
| **General terms** | 10 | 3 | | 13 |

## 6.2. Graded Relevance Judgment Categories

We developed our relevance scales as described below to categorize relevance judgment-decisions made by assessors, and accept the perception that the averaged judgment results of users will be taken as the "gold standard for performance evaluation" [25]. For each of the 30 ambiguous queries in Table 1, we define the corresponding information needs, which are assumed to be users' true information requirements. Human assessors are asked to decide, based on the defined information needs, which of the following four categories a returned Web snippet should belong to:

R:   relevant, the assessor is confident the link described by the Web snippet is relevant.

P:   partial relevant, the assessor believes the linked Web page may be relevant.

I:   irrelevant, the assessor is sure the link described by the Web snippet is irrelevant.

N:  the Web snippet doesn't provide sufficient information

Relevance judgment results from five different assessors are collected for each of the 30×50 = 1500 returned Web snippets. The assessors are PhD students from different areas and academic staff from our university. Since the queries are all commonly used general terms in daily life, thus no domain knowledge is needed to make a relevance judgment.

Each relevance judgment decision is assigned a numerical score, and a final score is calculated based on the scores. For the four defined judgments categories P, R, N and I, we assign 3, 1, 0, and -3 as the corresponding values. For each returned result, all assessors' relevance judgment scores will be added up to calculate a final score. A binary decision is reached based on the summarized score: the search results will be classified as relevant if the final score is positive, and as otherwise irrelevant. If the final score is zero, indicating no decision was made directly, we follow the link of the website, carefully review the full content of the linked webpage, and then decide if the website is relevant or irrelevant.

## 7. Experiments and Evaluation

### 7.1. Evaluation Measures

Precision, recall and P@10 are often used to measure the performance of IR systems [15]. We define a contingency table for each class to be evaluated in Table 3, where |TP| denotes the number of relevant items in the N returned results.

**Table 3.** Contingency table for category $i$.

| Category i | | True judgments | |
|---|---|---|---|
| | | **YES** | **NO** |
| **Classifier judgments** | **YES** | $TP_i$ (True Positive) | $FP_i$ (False Positive) |
| | **NO** | $FN_i$ (False Negative) | $TN_i$ (True Negative) |

With the contingency table, precision and recall are defined as

$$p_r = precision_i = \frac{TP_i}{TP_i + FP_i}$$

$$r_c = recall_i = \frac{TP_i}{TP_i + FN_i}$$

$$P@N = \frac{|TP|}{N}$$

### 7.2. Experimental Dataset Generation

There is a total of 50x30 = 1500 Web snippets collected from the meta-search engine used in the experiments. The dataset is generated by submitting the 30 queries listed in Table 1 to our meta-search engine to obtain the top 50 returned Web snippets. We have uploaded the returned Web snippets onto GitHub for research purpose, refer to our GitHub link in Section 4.

### 7.3. Human Relevant Judgment Results

After data collection, we employed 28 human judges with various skills to conduct relevant judgements. Judges are High Degree by Research students in the field of Accounting, Economics and Finance, Management, Marketing, and Information Systems from our university. The 28 judges are divided into six groups evenly (G1 to G6), with two assigned into two groups to ensure each group have five assessors. Each group is provided with 5x50 = 250 different Web search results from five different search terms. Assessors spent about 10 to 40 minutes finishing the relevancy judgment of the 250 Web snippets. Based on the value of the summarized scores (R=3, P=1, I=0, N=-3), we decide if a Web snippet is relevant ($s > 0$) or irrelevant ($s < 0$). If $s$ is zero, an assessor is asked to follow the links provided by the meta search engine to make a final relevant or irrelevant decision.

Following is an example of one relevance judgment result for the search term "resume" as presented in Tables 4 and 5. Table 4 shows Web snippets (*W-S*), four relevant judgment results (*R, P, I* and *N*); a final score (*SC*), true category (*RL*) and assessors' final judgment results (*JG*, a binary judgment as defined previously), new re-ranked results (*NR*), the judgment of the new results (*JG*), number of relevant documents in the new ranked results (*RL'*), recall of the re-ranked results (*Rc'*), and precision of the re-ranked results (*Pr'*). It also provides calculated precision as all the 50 results are reviewed at different recall levels (*Pr* and *Rc*). Table 5 summarizes the precision at ten different recall levels for the baseline search results and the re-ranked results (refer to next session for the re-ranked results).

### 7.4.    Re-ranking Search Results

The returned Web snippets from the meta-search engine are further processed via a re-ranking strategy which involves the following processes [33]:

1) Use the ODP data to create a training dataset where the ODP topics are taken as the labels of each item in the training dataset. Categories are Arts, Business, Computers, Games, Health, Home, News, Recreation, Reference, Regional, Science, Shopping, Society, Sports and Kids & Teens.
2) Use the generated training dataset to train a Naïve Bayes classifier to organize the 1500 Web snippets into different ODP topics listed above.
3) Use K-Nearest Neighbours further cluster the Web snippets into different clusters;
4) Merge the results from the above two steps.
5) Re-rank the results from step 4 based on the user preference profile which is assumed to contain two topics from the ODP topics aforementioned.

**Table 4.** Relevant judgment results of search term "resume" by five assessors [33]. The first column is shortened to save page space.

| W-S | R(3) | P(1) | I(-3) | N(0) | SC | JG | RL | Rc | Pr | NR | JG | RL' | Rc' | Pr' |
|---|---|---|---|---|---|---|---|---|---|---|---|---|---|---|
| 1. Resumes - | 1345 | | | 2 | 12 | 1 | 1 | 0.0303 | 1.0000 | 2 | 0 | 0 | 0.0000 | 0.0000 |
| 2. Résumé - | 5 | 4 | 13 | 2 | -2 | 0 | 0 | 0.0000 | 0.0000 | 4 | 1 | 1 | 0.0303 | 0.5000 |
| 3. Get. | 5 | 4 | 13 | 2 | -2 | 0 | 0 | 0.0000 | 0.0000 | 5 | 1 | 2 | 0.0606 | 0.6667 |
| 4. Resume | 345 | 1 | | 2 | 10 | 1 | 2 | 0.0606 | 0.5000 | 7 | 1 | 3 | 0.0909 | 0.7500 |
| 5. Resume | 1345 | | | 2 | 12 | 1 | 3 | 0.0909 | 0.6000 | 8 | 1 | 4 | 0.1212 | 0.8000 |
| 6. Entry | 1345 | | | 2 | 12 | 1 | 4 | 0.1212 | 0.6667 | 11 | 1 | 5 | 0.1515 | 0.8333 |
| 7. Free | 1345 | 1 | | 2 | 10 | 1 | 5 | 0.1515 | 0.7143 | 12 | 1 | 6 | 0.1818 | 0.8571 |
| 8. Resume - | 1345 | | | 2 | 12 | 1 | 6 | 0.1818 | 0.7500 | 13 | 1 | 7 | 0.2121 | 0.8750 |
| 9. JobStar: | 345 | 1 | | 2 | 10 | 1 | 7 | 0.2121 | 0.7778 | 22 | 1 | 8 | 0.2424 | 0.8889 |
| 10. Free | 1345 | | | 2 | 12 | 1 | 8 | 0.2424 | 0.8000 | 29 | 0 | 0 | 0.0000 | 0.0000 |
| 11. e-Resume | 5 | 34 | 1 | 2 | 2 | 1 | 9 | 0.2727 | 0.8182 | 38 | 0 | 0 | 0.0000 | 0.0000 |
| 12. e-resume. | 5 | 34 | 1 | 2 | 2 | 1 | 10 | 0.3030 | 0.8333 | 40 | 1 | 9 | 0.2727 | 0.7500 |
| 13. Professio | 35 | 4 | 1 | 2 | 4 | 1 | 11 | 0.3333 | 0.8462 | 14 | 0 | 0 | 0.0000 | 0.0000 |
| 14. resume: | | 5 | 134 | 2 | -8 | 0 | 0 | 0.0000 | 0.0000 | 17 | 0 | 0 | 0.0000 | 0.0000 |
| 15. Resume | 145 | | 3 | 2 | 6 | 1 | 12 | 0.3636 | 0.8000 | 19 | 1 | 10 | 0.3030 | 0.6667 |
| 16. e-Resume | | 45 | 13 | 2 | -4 | 0 | 0 | 0.0000 | 0.0000 | 20 | 1 | 11 | 0.3333 | 0.6875 |
| 17. FaxRe | | 5 | 1234 | | -11 | 0 | 0 | 0.0000 | 0.0000 | 25 | 1 | 12 | 0.3636 | 0.7059 |
| 18. Post your | | | 12345 | | -15 | 0 | 0 | 0.0000 | 0.0000 | 26 | 1 | 13 | 0.3939 | 0.7222 |
| 19. CV Res | 145 | | 3 | 2 | 6 | 1 | 13 | 0.3939 | 0.6842 | 27 | 1 | 14 | 0.4242 | 0.7368 |
| 20. Resume | 5 | 14 | 3 | 2 | 2 | 1 | 14 | 0.4242 | 0.7000 | 42 | 1 | 15 | 0.4545 | 0.7500 |
| 21. Resume | 35 | 4 | 1 | 2 | 4 | 1 | 15 | 0.4545 | 0.7143 | 49 | 0 | 0 | 0.0000 | 0.0000 |
| 22. Resume | 1 | | 345 | 2 | 6 | 1 | 16 | 0.4848 | 0.7273 | 1 | 1 | 16 | 0.4848 | 0.7273 |
| 23. Freshers | 345 | | 1 | 2 | 6 | 1 | 17 | 0.5152 | 0.7391 | 3 | 0 | 0 | 0.0000 | 0.0000 |
| 24. eResum | 345 | | 1 | 2 | 6 | 1 | 18 | 0.5455 | 0.7500 | 6 | 1 | 17 | 0.5152 | 0.7083 |
| 25. Resumes | 5 | 34 | 1 | 2 | 2 | 1 | 19 | 0.5758 | 0.7600 | 9 | 1 | 18 | 0.5455 | 0.7200 |
| 26. Sample | 135 | 4 | | 2 | 10 | 1 | 20 | 0.6061 | 0.7692 | 10 | 1 | 19 | 0.5758 | 0.7308 |
| 27. Resume | 145 | | 3 | 2 | 10 | 1 | 21 | 0.6364 | 0.7778 | 15 | 1 | 20 | 0.6061 | 0.7407 |
| 28. Careers | 5 | 14 | 3 | 2 | 2 | 1 | 22 | 0.6667 | 0.7857 | 16 | 0 | 0 | 0.0000 | 0.0000 |
| 29. Profes | 3 | | 145 | 2 | -6 | 0 | 0 | 0.0000 | 0.0000 | 18 | 0 | 0 | 0.0000 | 0.0000 |
| 30. Free | 1345 | | | 2 | 12 | 1 | 23 | 0.6970 | 0.7667 | 21 | 1 | 21 | 0.6364 | 0.7000 |
| 31. Resume | | 35 | 14 | 2 | -4 | 0 | 0 | 0.0000 | 0.0000 | 23 | 1 | 22 | 0.6667 | 0.7097 |
| 32. Resume | | 145 | 3 | | 2 | 10 | 1 | 24 | 0.7273 | 0.7500 | 24 | 1 | 23 | 0.6970 | 0.7188 |
| 33. Resumes | 5 | 34 | 1 | 2 | 2 | 1 | 25 | 0.7576 | 0.7576 | 28 | 1 | 24 | 0.7273 | 0.7273 |
| 34. Profess | | 5 | 134 | 2 | -8 | 0 | 0 | 0.0000 | 0.0000 | 30 | 1 | 25 | 0.7576 | 0.7353 |
| 35. Basic - | | 45 | 123 | | -7 | 0 | 0 | 0.0000 | 0.0000 | 31 | 0 | 0 | 0.0000 | 0.0000 |
| 36. Best | | 34 | 125 | | -7 | 0 | 0 | 0.0000 | 0.0000 | 32 | 1 | 26 | 0.7879 | 0.7222 |
| 37.Introduct | 15 | 34 | | 2 | 8 | 1 | 26 | 0.7879 | 0.7027 | 33 | 1 | 27 | 0.8182 | 0.7297 |
| 38. ESUME | | | 1345 | 2 | -12 | 0 | 0 | 0.0000 | 0.0000 | 34 | 0 | 0 | 0.0000 | 0.0000 |
| 39.The Write | 4 | | 135 | 2 | -6 | 0 | 0 | 0.0000 | 0.0000 | 35 | 0 | 0 | 0.0000 | 0.0000 |
| 40. What's | 135 | 4 | | 2 | 10 | 1 | 27 | 0.8182 | 0.6750 | 36 | 0 | 0 | 0.0000 | 0.0000 |
| 41. Resume | 1345 | | 2 | | 9 | 1 | 28 | 0.8485 | 0.6829 | 37 | 1 | 28 | 0.8485 | 0.6829 |
| 42. How to | 14 | 35 | | 2 | 8 | 1 | 29 | 0.8788 | 0.6905 | 39 | 0 | 0 | 0.0000 | 0.0000 |
| 43. Resume | 1345 | | | 2 | 12 | 1 | 30 | 0.9091 | 0.6977 | 41 | 1 | 29 | 0.8788 | 0.6744 |
| 44. Professio | | 3 | 145 | 2 | -8 | 0 | 0 | 0.0000 | 0.0000 | 43 | 1 | 30 | 0.9091 | 0.6818 |
| 45. Resume | | 14 | 3 | 25 | -1 | 0 | 0 | 0.0000 | 0.0000 | 44 | 0 | 0 | 0.0000 | 0.0000 |
| 46. Careers | | 345 | 1 | 2 | 1 | 1 | 31 | 0.9394 | 0.6739 | 45 | 0 | 0 | 0.0000 | 0.0000 |
| 47. resume | | 4 | 13 | 25 | -5 | 0 | 0 | 0.0000 | 0.0000 | 46 | 1 | 31 | 0.9394 | 0.6596 |
| 48. Resume | 1345 | | | 2 | 12 | 1 | 32 | 0.9697 | 0.6667 | 47 | 0 | 0 | 0.0000 | 0.0000 |
| 49. Create a | | 4 | 13 | 25 | -5 | 0 | 0 | 0.0000 | 0.0000 | 48 | 1 | 32 | 0.9697 | 0.6531 |
| 50. How To | 1345 | | | 2 | 12 | 1 | 33 | 1.0000 | 0.6600 | 50 | 1 | 33 | 1.0000 | 0.6600 |

**Table 5.** Precision at different recall levels for search term "resume" [33]

| Rc-Lv | 10 | 20 | 30 | 40 | 50 | 60 | 70 | 80 | 90 | 100 |
|---|---|---|---|---|---|---|---|---|---|---|
| Pr meta | 83.3 | 83.3 | 83.3 | 76.9 | 76.9 | 76.9 | 76.7 | 69.8 | 69.8 | 66 |
| Pr Re-Ranked | 87.5 | 87.5 | 74.1 | 74.1 | 74.1 | 74.1 | 73 | 73 | 68.2 | 66 |

The evaluation results are presented in Fig. 1, which contains precision-recall curve of the baseline results from the meta-search engine, and the re-ranked results based on the above process. Note that the curve is drawn based on the averaged results over the 30 search terms. The relevance judgment outcomes are the summed-up results of five judges for each search terms.

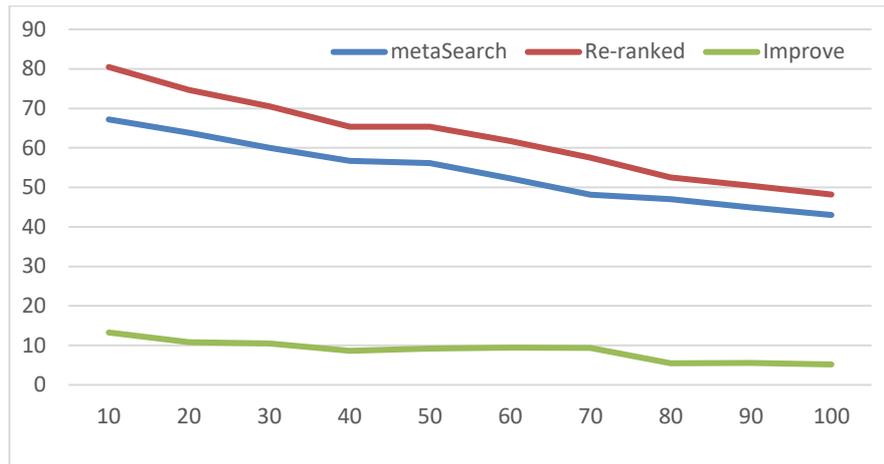

**Fig 1.** Precision-recall curve of meta-search results and re-ranked results

Fig. 1 illustrates that:

- The re-ranked results outperform the meta-search engine results consistently on all recall level. The maximum improvement is 12.06% at the recall level of 10%, and the minimum improvement is 5.18% at the recall level of 100%.
- The averaged precisions over all 30 queries of meta-search engine and re-ranked results are 55.55% and 64.29% respectively; this indicates an average 8.74% precision improvement.

The improvements of re-ranked results over baseline meta-search engine results decreases as recall increases; the maximum increase happened at recall level 10%, and it drops down the way to a minimum as recall increases to 100%. This is a preferable outcome as users usually browse only a few pages of Web search results, and about 50% of them only browse the first page [12].

### 7.5. Relevance Judgment Convergence Degree (RJCD)

While the average performance of re-ranked search results consistently exhibits superior performance to the baseline meta-search engine results, we also observed that there are nine search terms (namely maps, music, jokes, games, Disney, resume, Susan Dumais, graphic design, and Saturn) for which the baseline results outperform the re-ranked results marginally. To further investigate the situation, the concept of RJCD is introduced to depict for a given search term, to what degree the relevancy assessors agree with each other; and thus propose to use RJCD as a criteria to measure the quality of a query.

Let $h$ be the number of human assessors, and $k$ be the number of relevance judgment options an assessor can select from, here $k = 4$ corresponding to the four options $R$, $P$, $N$ and $I$. If $N$ is the total number of Web snippets returned by an IR system for a given query, we define *Agreement Number AN* = the total number the sort of judgments for that all $h$ judges to make the same relevance judgment decision; and *Judgment Number JN* = the total number of choices made by the $h$ judges over all $N \times k$ possible choices. Formally, we denote $\Omega = \{R, P, N, I\}$, $k = |\Omega|$, a relevance judgment by assessor $j$ for the $i$-th returned result as $R_j(i) \in \Omega$, $j \in [1, \dots h]$, $i \in [1, \dots N]$, further, let

$$\gamma(i) = \left\| \bigsqcup_{j=1}^{h} R_j(i) \right\| = |\Omega'|$$

be the size of Ω' ⊆ Ω which contains distinct relevance judgment results from Ω. We specially define γ(i)|₁ ≡ 1, which indicates that all assessors reach the same judgment decision, no matter what the relevant result it is; for example, all h judges give the *R* decision. RJCD for a given query can then be defined in the following equation as ρ

$$\rho = \frac{\sum_{i=1}^{N} \gamma(i)|_1}{\sum_{i=1}^{N} r(i)} \triangleq \frac{AN}{JN}$$

Use data in Table 4 as an example, for search term "resume", we need first calculate γ(1), γ(2), … γ(50). By definition, γ(1) = |{R, N}|=2, γ(2) = |{R, P, I, N}| = 4,…, γ(18) = |{I}| = 1,…, so we can get JN by summing up γ(i). Note also that we have only γ(18) = 1, that is, *AN* = 1, so our final score of RJCD is 1/145=0.006897.

Fig. 2 illustrates the relationship between RJCD and the precision improvement of the re-ranked results over the baseline meta-search engine results; it demonstrates that there are positive relations between precision improvement and the values of the RJCD. We will analyse its correlation in the following session.

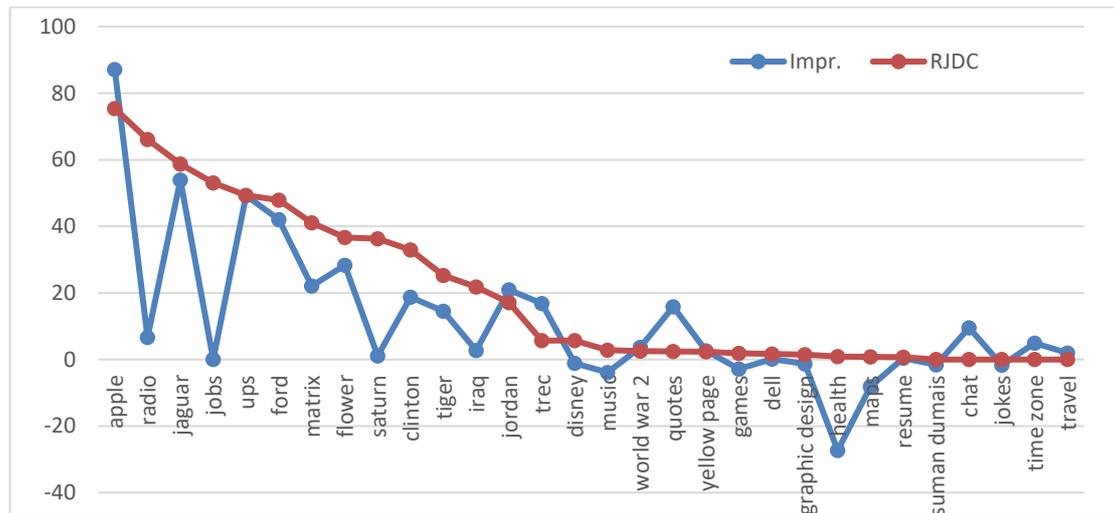

**Fig. 2.** Precision improvement of the re-ranked results and RJDC values over the 30 queries

## 7.6. Correlation Analysis between RJCD and the Re-ranked Results

Fig. 2 demonstrates an improvement of precision attribute with a corresponding increasement in RJCD. When RJCD is small, the corresponding precision improvements are either very small or even negative; indicating the re-ranked results are worse than the baseline meta search engine results. The average RJCD score is only 5.5% for the nine negative search terms; while for the positive queries, the average RJCD score is about 20%.

Let us estimate the correlation coefficient of precision improvement and RJCD. The correlation coefficient is defined for two random variables X and Y [18]:

$$\beta = \frac{cov(X,Y)}{\sqrt{V(X)V(Y)}} = \frac{\sigma_{XY}}{\sigma_X \sigma_Y} = \frac{E[(X - \mu_X)(Y - \mu_Y)]}{\sigma_X \sigma_Y}$$

where cov(X,Y), also denoted as σ_XY, is the covariance of X and Y, μ_X is the mean value of X, μ_Y is the mean value of Y, V(X), V(Y) are the variance of X and Y, which are denoted as $\sigma_X^2$ and $\sigma_Y^2$ defined as

$$\sigma_X^2 = V(X) = E(X - \mu)^2$$
$$\sigma_Y^2 = V(Y) = E(Y - \mu)^2$$

The correlation coefficient computed between RJCD and precision improvement is 0.725, with p-value 0.000006, which strongly indicates that the two variables are positively related. It can also be observed by the trends of RJCD in Fig 2, where when RJCD is high (on the left of Fig 2.), the precision improvements are also high; and as the RJCD reduces to zero in the right part of Fig 2, the improvements are marginal or negative. Therefore, RJCD is a reliable measurement to evaluate if a search term is a good representation of users' real information needs or not. If RJCD is less than 5%, we recommend it is reasonable to use an alternative search term to represent the users' information needs, and the search term with low RJCD should not be included in a labelled training dataset to be used to evaluate the performance of IR and related systems.

## 8. Conclusion and Future Work

Relevancy judgment is an essential part of evaluating IR systems. Previous research focuses more on the agreement among different assessors or assessor groups, where full length documents are available. For Web IR systems, users need to make a relevance judgment decision based on the returned search results, or Web snippets which are usually much less informative than normal documents. When a dataset is created to evaluate the performance of a Web IR system, we suggested the quality of queries should also be measured by the proposed RJCD to exclude those that are too ambiguous to make relevance judgments of assessors largely diverge. Relevance Judgment Convergence Degree was employed in the research as a criterion to measure the quality of ambiguous queries in Web IR evaluation and test datasets construction. We evaluated the performance of a baseline IR system based on a meta-search search engine with 30 ambiguous search terms and top 50 Web snippets for each of the queries. We then improved the ranking of the baseline results by combining classification and clustering techniques. Experimental results revealed positive correlation coefficient exists between RJCD and performance improvements. We recommended that if the RJCD of a query is less than 5%, the query and the returned search results should not be included in the test dataset.

In future, we will extend our experiments with more queries and Web IR systems to verify the effectiveness of RJCD as a criterion to measure the quality of ambiguous search terms. Meanwhile, we will validate RJCD effectiveness by examining correlation coefficient among RJCD and the performance improvements among Web IR systems.